\documentclass[10pt,twocolumn,letterpaper]{article}

\usepackage[final, pagenumbers]{cvpr}      
\usepackage{capt-of}
\usepackage{adjustbox}
\definecolor{cvprblue}{rgb}{0.21,0.49,0.74}
\usepackage{multirow}
\usepackage[pagebackref,breaklinks,colorlinks,allcolors=cvprblue]{hyperref}

\def\paperID{4047} 
\def\confName{CVPR}
\def\confYear{2026}

\usepackage{booktabs}
\usepackage[table]{xcolor}
\definecolor{lightblue}{RGB}{235,245,255}

\title{SounDiT: Geo-Contextual Soundscape-to-Landscape Generation}

\author{%
    Junbo Wang$^{1,2,\dagger}$ \quad
    Haofeng Tan$^{1,3,\dagger}$ \quad
    Bowen Liao$^{4}$\quad
    Albert Jiang$^{1}$\quad
    {Teng Fei}$^{5}$\quad \\
    {Qixing Huang}$^{1}$\quad
    {Bing Zhou}$^{2}$ \quad
    {Zhengzhong Tu}$^{6}$\quad 
    {Shan Ye}$^{7,\ddagger}$\quad
    {Yuhao Kang}$^{1,\ddagger}$\\
    $^1$The University of Texas at Austin \quad \quad
    $^2$University of Tennessee, Knoxville \\
    $^3$University of South Carolina\quad
    $^4$Arizona State University\quad
    $^5$University of Canterbury\\
    $^6$Texas A\&M University \quad\quad\quad
    $^7$University of Wisconsin-Madison\\
    \small{$^\dagger {^*} $Equal contributions. $^\ddagger$Corresponding Author: Yuhao Kang (yuhao.kang@austin.utexas.edu).}
}

\begin{document}
\twocolumn[{%
\maketitle
\vspace{-0.4em}
\vspace{-10mm}
\begin{center}
  \includegraphics[width=\textwidth]{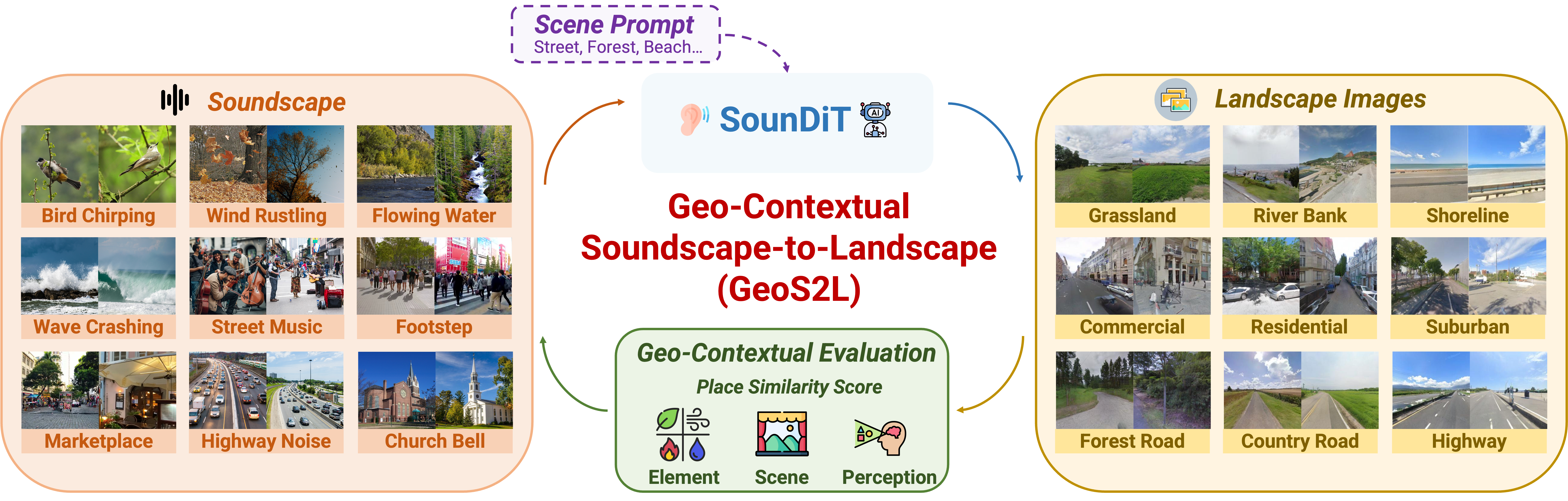}
  \captionof{figure}{\textbf{Geo-contextual soundscape-to-landscape (GeoS2L) generation}
  aims to synthesize realistic landscape images from environmental soundscapes. We introduce large-scale geo-contextual datasets, the SounDiT model, and the Place Similarity Score evaluation framework to support this task.}
  \label{fig:teaser}
\end{center}
\vspace{0.6em}
}]

\begin{abstract}

Recent audio-to-image models have shown impressive performance in generating images of specific objects conditioned on their corresponding sounds. However, these models fail to reconstruct real-world landscapes conditioned on environmental soundscapes. To address this gap, we present Geo-contextual Soundscape-to-Landscape (GeoS2L) generation, 
a novel and practically significant task that aims to synthesize geographically realistic landscape images from environmental soundscapes. To support this task, we construct two large-scale geo-contextual multi-modal datasets, SoundingSVI and SonicUrban, which pair diverse environmental soundscapes with real-world landscape images. We propose SounDiT, a diffusion transformer (DiT)-based model that incorporates environmental soundscapes and geo-contextual scene conditioning to synthesize geographically coherent landscape images. Furthermore, we propose the Place Similarity Score (PSS), a practically-informed geo-contextual evaluation framework to measure consistency between input soundscapes and generated landscape images. Extensive experiments demonstrate that SounDiT outperforms existing baselines in the GeoS2L, while the PSS effectively captures multi-level generation consistency across element, scene,and human perception.  Project page: \url{https://gisense.github.io/SounDiT-Page/}




\end{abstract}    
\section{Introduction}
\label{sec:intro}
\noindent\textit{``The earth has music for those who listen.''}\par
\hfill — William Shakespeare

Environmental soundscapes encode rich geographic information, often in greater detail than the specific object acting as the sound source \citep{ruiz2024urban, berleant1997living, blesser2007spaces, algargoosh2022impact}. 
A burst of birdsong, for example, implies a leafy trail or an urban green space, while the sound of traffic may evoke a busy streetscape, beyond a single car.
Recognizing this strong geo-contextual connection between auditory soundscape and visual landscape, researchers in geography \citep{tuan1975images,cosgrove2012geography,krygier1994sound}, urban planning \citep{de2010soundscape, kang2023soundscape,hemmat2023exploring}, and environmental psychology \citep{mehrabian1974approach,guastavino2007categorization}, have long studied the impacts of soundscapes. 
For example, how analyzing specific sound types can inform and improve urban design\cite{jeon2012acoustical}, how to mitigate noise through environmental design \cite{brown2004approach, brown2014soundscape}, or which soundscapes are associated with human restoration and well-being \cite{zhang2017effects, uebel2021urban}.
Previous practices in these fields have often been limited to descriptive statistics of the soundscape environment (e.g., decibel levels, sound types) \cite{jo2021urban, chen2023natural}, which are not always intuitive and fail to capture the co-located visual characteristics of the scene.
The recent emergence of deep generative models for Audio-to-Image (A2I) synthesis \cite{sung2023sound, qin2023gluegen, yariv2023audiotoken, yang2024visual, biner2024sonicdiffusion}, which translates auditory information directly into visual representations, offers a promising opportunity to model the geo-contextual soundscape-landscape relationships, unlocking significant practical applications.

However, we observe three noticeable gaps when applying prior A2I in geographic contexts and planning practices for associating soundscapes and landscapes.
First, most A2I studies rely on generic audio-visual datasets such as object audio \cite{owens2016visually,zhou2018visual,chen2020vggsound}, human voice \cite{nagrani2017voxceleb, livingstone2018ryerson}, weathers \cite{li2022learning}, and limited scene types \cite{lee2022sound}.
Consequently, these models link an audio signal to its source (e.g., generating an image of a bird for a bird call) rather than its environment, overlooking geo-contextual insights vital for real-world applications, such as parks vs. beaches and urban vs. rural settings \cite{yuan2012discovering}. 
For geographers and urban planners, the primary interest is not the object making the sound (that bird, that car), but the environmental setting that informs environmental design and noise mitigation strategies. Despite explorations of diffusion-based models in geographic audio-to-image settings \citep{zhuang2024hearing,zhuang2025soundscape}, these methods still take soundscapes as the sole input and output landscape images, without geographic context.
Second, Diffusion Transformer (DiT) \cite{peebles2023scalable} techniques have recently demonstrated impressive results across various tasks such as image \cite{li2024hunyuan, chen2023pixart, chen2024pixart} and video generations \cite{ma2024latte, wang2024av, kong2024hunyuanvideo, liu2024sora}. 
However, they remain unexplored in A2I generation and have yet to incorporate geographic knowledge to support applications in geography and urban planning.
Furthermore, common evaluation metrics employed in prior studies, such as the Fréchet Inception Distance (FID) \cite{heusel2017gans}, may fall short in capturing geographic semantic alignment between the input soundscape and generated images. 
Ideally, the evaluation metrics should consider environmental characteristics and geographic contexts: 
the place setting depicted in the input soundscape and those represented in the generated images should be similar to ensure geographic contextual coherence and relevance.


Given these challenges, we introduce and formalize a new, practically-motivated task: Geo-Contextual Soundscape-to-Landscape (GeoS2L) generation (Figure \ref{fig:teaser}), 
to synthesize realistic landscape images from environmental soundscapes, optionally conditioned on scene context. 
The GeoS2L problem extends traditional A2I synthesis by incorporating geo-contextual insights with significant practical implications. 
Here, \textit{landscape} refers to 
geographical environments shaped by natural and built features \cite{10.7551/mitpress/9780262015523.001.0001}.
A \textit{soundscape} denotes the acoustic environment as perceived by humans at places, comprising sounds from both natural and anthropogenic sources (e.g., bird calls and vehicle noises) \cite{Schafer1977-SCHTTO-12}.
A \textit{scene} indicates place types such as beach, forest, street, to guide image generation. Our contributions span three key areas: 
\underline{First}, we introduce two large-scale, multi-modal geo-contextual datasets: \textit{SoundingSVI} and \textit{SonicUrban}, composed of over 169K and 237K soundscape-landscape image pairs. These datasets cover diverse geographic locations and scenes, supporting geo-contextual realistic synthesis.
\underline{Second}, we develop \textit{SounDiT}, an innovative generative model that support GeoS2L problem.
It introduces three novel, efficient components to integrate geo-contextual information: a MoE Soundscape Conditioning module, a Scene Low-Rank Content Mixer (SLRCM), and a Scene AdaLn (S-AdaLn). These modules effectively inject geographic contexts to environmental soundscapes, enabling the generation of geographically coherent landscape images.
\underline{Third}, we design a novel practically-motivated geo-contextual evaluation framework: \textit{Place Similarity Scores (PSS)}. PSS includes three metrics: element-level, scene-level, and perception-level similarity
to assess geographic contexts and environmental consistency of generated landscape images. 
Extensive experiments demonstrate that our model achieves state-of-the-art performance and delivers strong utility on downstream applications like geography, environment, and urban planning with significant practical impacts. Moreover, our datasets and evaluation metric provide a robust benchmark for the GeoS2L task.

\section{Related Work} 
\noindent\textbf{Audio-to-Image Synthesis} aims to translate acoustic features into coherent and meaningful images and has gained increasing attention with the advancement of various multi-modal alignment foundation models \cite{yang2024visual,sung2023sound} and generative techniques \cite{goodfellow2014generative, rombach2022high}. Current A2I models are broadly divided into GAN-based (e.g., sound2scene \cite{sung2023sound}) and Diffusion-based (e.g., AudioToken \cite{yariv2023audiotoken}, and GluenGen \cite{qin2023gluegen}). These models mainly rely on generic datasets which lack geo-contextual information, often limited to object audios \cite{owens2016visually}, human voice \cite{nagrani2017voxceleb}, weather \cite{li2022learning}, and broad scene types \cite{lee2022sound}. As a result, they often produce stylized or unrealistic outputs and fail to capture the geographic coherence between soundscapes and landscapes.

\noindent\textbf{Landscapes and Soundscapes} represent two dimensions for characterizing places \cite{liu_landscape_2013, fuller_connecting_2015}.
Landscapes offer visual representations of the physical layout and built environments, and are
commonly captured via street view imagery (SVI) and remote sensing (RS) images \cite{lee2022sound,li2022learning}.
In parallel, soundscapes refer to the acoustic environment perceived at specific places.
Existing datasets, such as
BirdCLEF \cite{kahl-2019-overview}, BirdSet \cite{rauch_birdset_2024},
SoundingEarth \cite{heidler2023self}), AudioSet \cite{gemmeke_audio_2017}, UrbanSound8K \cite{salamon_dataset_2014}, SONYC \cite{cartwright_sonyc-ust-v2_2020}, EnigenScape \cite{green_eigenscape_2017}, and EMO-Soundscape \cite{fan_emo-soundscapes_2017},
have been utilized to support real-world practices including monitoring biodiversity \cite{rasmussen2024sound}, noise mitigation \cite{huang2024estimating}, and human perception analysis \cite{fang2021soundscape}.
However, prior A2I studies have overlooked such geographic and environmental insights, limiting their 
real-world applications.
To address this gap, our work integrates geo-contextual information for multimodal generative modeling by constructing datasets and evaluation frameworks.


\begin{figure*}[t]
  \centering
  \includegraphics[width=\textwidth]{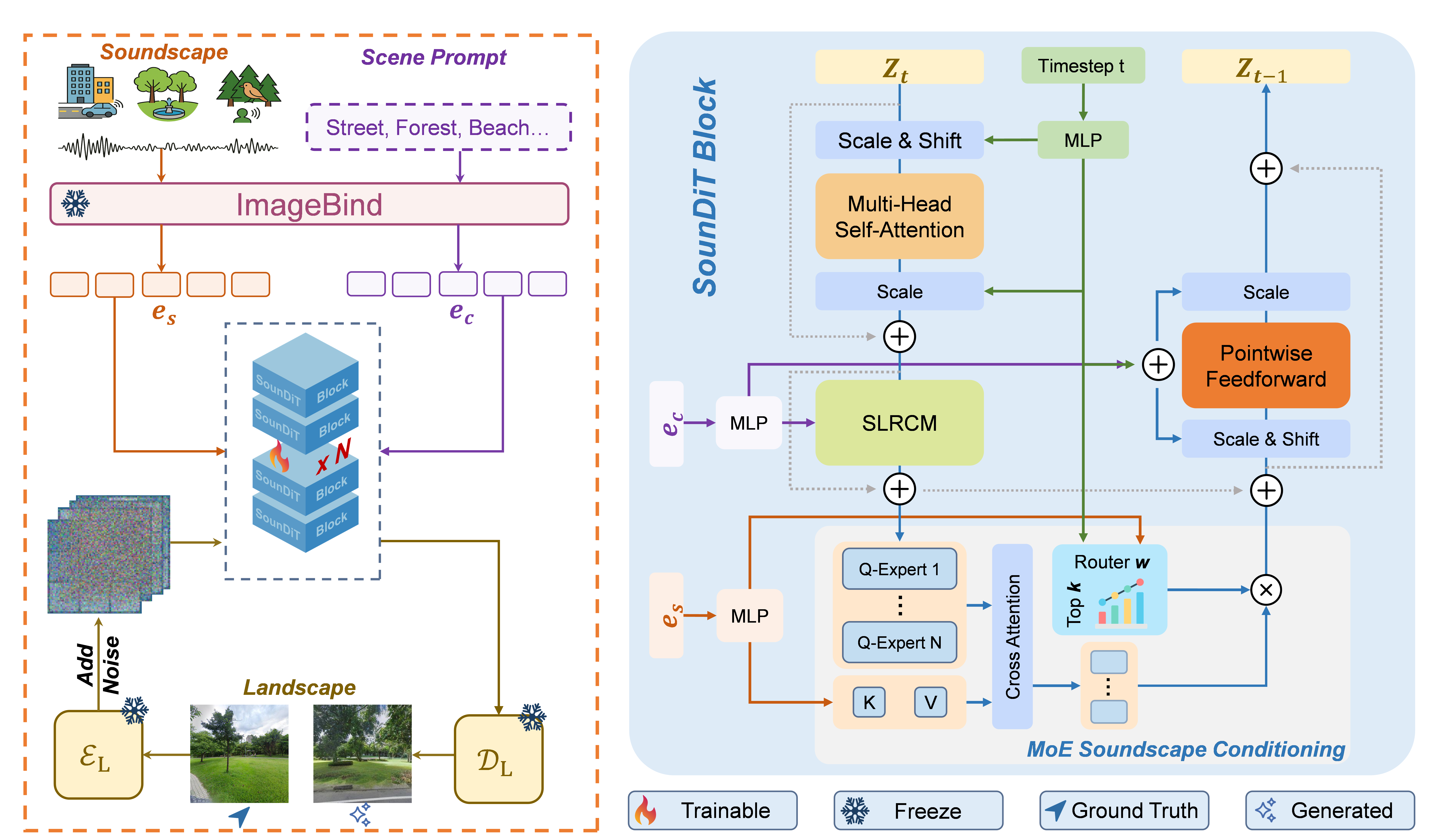}
    \caption{\textbf{The \textit{SounDiT} framework.}
    \textit{SounDiT} encodes soundscape and scene inputs into a shared space with pre-trained encoders, and performs denoising in latent space with DiT blocks equipped with AdaLN-Zero timestep conditioning. Scene information is injected through a Scene Low-Rank Mixer (SLRCM) and a lightweight Scene AdaLN (S-AdaLN) path to enhance geo-contextual consistency, while a Mixture-of-Experts (MoE) Soundscape conditioning module aligns soundscape features with visual landscape tokens.}
    
  \label{fig:pipeline}
\end{figure*}

\section{Methodology} 
\subsection{Preliminary}
\noindent\textbf{Diffusion Model.} We adopt a latent-based diffusion model, consisting of a forward diffusion and backward denoising process in the latent space \(\mathbf{z}\). The forward diffusion \(\mathbf{q}\) starts from the initial latent \(\mathbf{z}_0\), gradually adding Gaussian noise over timestep \(T\) based on a fixed variance $\alpha$ and defined as:
\begin{equation}
q(\mathbf{z}_{t} \mid \mathbf{z}_{t-1}) = \mathcal{N}\bigl(\mathbf{z}_{t}; \sqrt{\alpha_t}\, \mathbf{z}_{t-1}, (1 - \alpha_t)\, \mathbf{I} \bigr).
\end{equation}
The backward denoising process \(p_\theta\) is trained to predict the less noisy latent:
\begin{equation}
p_{\theta}(\mathbf{z}_{t-1} \mid \mathbf{z}_{t}) = \mathcal{N} \bigl( \mathbf{z}_{t-1}; \, \mu_{\theta}(\mathbf{z}_t, t, v), \, \Sigma_{\theta}(\mathbf{z}_t, t, v) \bigr).
\end{equation}
where  \(v\) is the conditional signal and \(\mu_{\theta}\), \(\Sigma_{\theta}\) are predicted through a denoising network \(\epsilon_{\theta}(\mathbf{z}_t, t, v)\).

\noindent\textbf{Classifier-Free Guidance.} 
To control the influence of the conditioning signal \(v\) during backward denoising, we employ classifier-free guidance (CFG) to predict noise \(\hat{\boldsymbol{\epsilon}}\) based on both conditional and unconditional signals with specific guidance scale \(s\).
\begin{equation}
\hat{\boldsymbol{\epsilon}} = \boldsymbol{\epsilon}_{\theta}(\mathbf{z}_t, t, \varnothing) + s \bigl[ \boldsymbol{\epsilon}_{\theta}(\mathbf{z}_t, t, v) - \boldsymbol{\epsilon}_{\theta}(\mathbf{z}_t, t, \varnothing) \bigr].
\end{equation}
\subsection{GeoS2L Task Formulation}
We formulate our task as \emph{Geo-Contextual Soundscape-to-Landscape (GeoS2L)}, extending A2I synthesis with geographic contexts.
GeoS2L aims to synthesize landscape images from environmental soundscapes guided by an optional scene prompt. 
We define a \emph{landscape} $l$ as the geographical–ecological environment shaped by natural and built features, a \emph{soundscape} $s$ as the human-perceived acoustic environment at a place, including natural and anthropogenic sources (e.g., bird calls and traffic noise).
However, real-world soundscape recordings often lack sufficient information to fully capture the visual characteristics of the surrounding environment. 
For example, a bird call may occur in either rural or urban parks.
To provide greater flexibility and user control, we incorporate an optional \emph{scene prompt} $c$ as the semantic geographic context (e.g., park, beach, street).
This prompt serves as an additional geo-contextual conditioning signal to facilitate generative process toward a specific, context-aware landscape.
Given a dataset
\(
D=\{(s_i,c_i,l_i)\mid i=1,\dots,N\},
\)
our goal is to learn a generator
\(
G:\,(s_i,c_i)\mapsto \hat{l}_i
\)
such that the synthesized image $\hat{l}_i$ is both visually realistic and geo-contextually consistent with the ground truth $l_i$, quantified by the relevance function $R(s,c,l)$. This can formulated as
\begin{equation}
\mathcal{L}\;=\;\mathbb{E}_{(s_i,c_i,l_i)\sim D}\!\left[\,R(s_i,c_i,l_i)\;-\;R\bigl(s_i,c_i,\hat{l}_i\bigr)\right].
\label{eq:concept}
\end{equation}
This enables the model to capture cross-modal correlations among acoustic, semantic, and visual cues, producing landscape images that reflect both the soundscape and the intended scene context.


\subsection{SounDiT}
To effectively and efficiently inject soundscape and scene context, each SounDiT block introduces (1) a Mixture-of-Experts (MoE) Soundscape Conditioning module, (2) a Scene Low-Rank Mixer (SLRCM) module, and a Scene AdaLN (S-AdaLN) module to improve scene consistency. 

\noindent\textbf{Multimodal Encoders.}
Following latent diffusion models (LDM) \cite{rombach2022high}, landscape images are reconstructed within the latent space.
Given a landscape image $l\!\in\!\mathbb{R}^{H\times W\times 3}$, a VAE encoder–decoder $(\mathcal{E}_L,\mathcal{D}_L)$ is employed to encode it into the latent embedding \(
\mathbf{e}_l = \mathcal{E}_L(l)\in\mathbb{R}^{h\times w\times c}
\) and reconstruct the latent embedding back into a image \(
\hat{l} = \mathcal{D}_L(\mathbf{e}_l).
\label{eq:vae}
\)
To integrate acoustic features and geo-contextual semantics into denoising process, a multi-modal encoder is adopted to encode the soundscapes $s$ and the scene prompt $c$ into a shared latent space, resulting in a soundscape embedding $\mathbf{e}_s$ and a scene embedding $\mathbf{e}_c$.

\noindent\textbf{SounDiT block.}
The SounDiT block operates in four stages. First, we apply multi-head self-attention layer conditioned by the timestep embedding \(e_{\mathrm{t}}\) via AdaLN-Zero \cite{goyal2017accurate}, preserving compatibility with pretrained DiT backbones.
Second, we design a lightweight \textbf{Scene Low Rank Context Mixer (SLRCM)} module to inject scene context.
SLRCM introduces a low-rank residual path inside each SounDiT block.
Given an input token $x\in\mathbb{R}^{B\times N\times D}$ and scene embedding $\mathbf{e}_c\in\mathbb{R}^{B\times D_c}$, SLRCM modulates tokens through a rank-$r$ linear operator parameterized by $\mathbf{e}_c$. 
We specify the low-rank content mixer \(A(\mathbf{e}_c)\) that transforms token features along an \(r\)-dimensional path, together with a sample-wise strength \(s(\mathbf{e}_c)\) that controls the overall contribution of this path:
\begin{equation}
A(\mathbf{e}_c)=W_q\,\mathrm{Diag}\!\big(\tanh(\phi(\mathbf{e}_c))\big)\,W_v,
\label{eq: scene-injection3}
\end{equation}
where \(W_q\!\in\!\mathbb{R}^{D\times r}\) and \(W_v\!\in\!\mathbb{R}^{r\times D}\) are learned low-rank projections. 
The mapping \(\phi:\mathbb{R}^{D_c}\!\to\!\mathbb{R}^r\) converts the scene embedding into an \(r\)-dimensional gating vector, and the diagonal operator applies this gate element-wise along the rank-\(r\) channel, yielding a compact, diagonally gated low-rank transform. In addition, we define a positive, sample-wise strength to regulate the residual magnitude:
\begin{equation}
s(\mathbf{e}_c)=g(\mathbf{e}_c)\,\mu\,\tanh(\alpha),
\label{eq: scene-injection2}
\end{equation}
where \(g:\mathbb{R}^{D_c}\!\to\!\mathbb{R}_{>0}\) produces a per-sample scale (e.g., via softplus), \(\mu\) is a global guidance scalar, and \(\alpha\) is a learnable scalar bounded. Given these definitions, the token update becomes:
\begin{equation}
x' \;=\; x \;+\; s(\mathbf{e}_c)\,\mathrm{LN}(x)\,A(\mathbf{e}_c),
\label{eq: scene-injection}
\end{equation}
where \(\mathrm{LN}(\cdot)\) denotes layer normalization. SLRCM thus provides a compact, diagonally gated low-rank modulation that preserves the pre-trained attention structure while injecting scene priors with low computational costs. 

Third, to further integrate multi-level soundscape features, we introduce \textbf{MoE Soundscape Conditioning} module, composed of \(M\) experts that share the key $K_s$ and value $V_s$ while maintaining expert-specific, low-rank query $Q$.
Given input token $x\in\mathbb{R}^{B\times N\times D}$,we compute \(K_s = s\,W_K, V_s = s\,W_V\) once and for each expert $m\in\{1,\dots,M\}$ form a low-rank query \(Q_m\)
\begin{equation}
Q_m = f(x)\,W^{(m)}_{q\downarrow} W^{(m)}_{q\uparrow}\!,
\label{eq:share KV-ExpertQ}
\end{equation}
where $f$ is a per-token Layer Norm. Each expert then produces
\begin{equation}
Z_m \;=\; \mathrm{MHA}\bigl(Q_m,\,K_s,\,V_s\bigr)\,W_O,
\label{eq:share KV-ExpertQ-Cross}
\end{equation}
where $W_O$ is shared across experts, $Z_m$ refers to soundscape-conditioned token features, and $MHA$ is multi-head attention.

Routing weights are computed by a temperature-scaled dot product between the audio summary and learnable prototypes $\mathbf{P}\in\mathbb{R}^{d\times M}$ with time augmentation introduced through a projected timestep embedding:
\begin{equation}
w
=
\operatorname{Softmax}\!\Bigl(
  \tfrac{1}{\tau}\;\mathbf{e}_s^\top
  \bigl(\mathbf{P} + \mathbf{W}_t\,e_t\,\mathbf{1}^\top\bigr)
\Bigr),
\label{eq:audio-injection1}
\end{equation}
where $e_t$ is the timestep embedding projected by $\mathbf{W}_t$.
Finally, we aggregate the experts via a top-$k$ soft mixture and apply a global audio gate to update tokens in each SounDiT block:
\begin{equation}
x' \;=\; x \;+\; \gamma\sum_{m\in\mathcal{K}} 
\operatorname{softmax}\!\bigl(w_{m}/\tau_m\bigr)_m \; Z_m,
\label{eq:audio-update}
\end{equation}
where $\gamma = \tanh(\mathbf{e}_s)$ is a scalar gate, yielding a smooth, zero-centered bounded activation.
By enabling expert specialization under a fixed K/V budget, the MoE provides scalable conditioning that improves geo-contextual consistency across diverse soundscapes.

Finally, to enhance geo-contextual consistency, we introduce a practical-informed \textbf{S-AdaLN}, an AdaLN-Zero extension, which derives scale–shift parameters from the timestep embedding $\mathbf{e}_t$ and the scene embedding $\mathbf{e}_c$ via learnable bounded mixing. 
The modulated tokens are processed by a pointwise feed-forward network, followed by a gated residual add, yielding the block output for noise-residual prediction. By hierarchically fusing visual, scene, and soundscape cues, SounDiT generates geo-contextually coherent landscapes.

\section{Datasets} 
\begin{table}[t]
  \centering
  \footnotesize
  \setlength{\tabcolsep}{3pt}
  \renewcommand{\arraystretch}{1.1}
  \caption{\textbf{Comparison of audio-visual datasets.} \(\dag\) denotes partial/limited geo-context.}
  \resizebox{\columnwidth}{!}{%
  \begin{tabular}{l|cccc}
    \toprule
    \textbf{Dataset} & \textbf{Modalities} & \textbf{\# Size} & \textbf{Geo-Contextual} & \textbf{Viewpoint} \\
    \midrule
    ACAV100M \cite{lee2021acav100m}     & Video            & 100M  & \(-\)          & Ground \\
    \hline
    VEGAS \cite{zhou2018visual}          & Video            & 28K   & \(-\)          & Ground \\
    \hline
    VGGSound \cite{chen2020vggsound}     & Video            & 200K  & \(\dag\)       & Ground \\
    \hline
    IntoTheWild \cite{li2022learning}    & Video            & 142   & \checkmark     & Ground \\
    \hline
    Landscape \cite{lee2022sound}        & Video            & 1747  & \checkmark     & Ground \\
    \hline
    SoundingEarth \cite{heidler2023self} & Audio \& Image   & 50K   & \checkmark     & Overhead \\
    \midrule
    \cellcolor{lightblue}\textbf{SoundingSVI (Ours)} &
      \cellcolor{lightblue}Audio \& Image & \cellcolor{lightblue}169K &
      \cellcolor{lightblue}\checkmark & \cellcolor{lightblue}Ground \\
    \hline
    \cellcolor{lightblue}\textbf{SonicUrban (Ours)} &
      \cellcolor{lightblue}Video & \cellcolor{lightblue}237K &
      \cellcolor{lightblue}\checkmark & \cellcolor{lightblue}Ground \\
    \bottomrule
  \end{tabular}
  }
  \label{tab:av-datasets}
\end{table}

Prior A2I studies have primarily relied on large-scale, general-domain audio-visual datasets, such as VEGAS \cite{zhou2018visual}, VGGSound \cite{chen2020vggsound}, and ACAV100M \cite{lee2021acav100m}. These datasets primarily focus on sound-source localization and general audio-visual correspondence learning, lacking the specific geographical context required for soundscape-to-landscape generation.
For instance, geographers may be more interested in generating a wetland implied by a frog call, rather than an image of the frog itself. 
A car horn might suggest a congested urban street, rather than a static vehicle.
Some datasets, such as Landscape \cite{lee2022sound} and IntoTheWild \cite{li2022learning}, have included paired soundscape and image data, but offer limited spatial coverage and scene diversity.
To address this gap, we introduce two new multimodal datasets: \textit{SoundingSVI} and \textit{SonicUrban},
designed to support GeoS2L generation with large-scale, geographically diverse, and context-rich data.

\noindent\textbf{SoundingSVI.} The SoundingSVI dataset was constructed from geotagged soundscape recordings from the Aporee platform \cite{noll2013radio}.
For each raw soundscape recording, we retrieved nearby Google Street View Image (SVI) using its latitude and longitude to capture the surrounding visual landscapes.
This follows a similar approach to the \textit{Sounding Earth} dataset \cite{heidler2023self}, which paired geotagged audio with remote sensing imagery. The raw recordings were then segmented into 10-second clips, with a human voice detection model \cite{silero2024vad} used to filter clips dominated by human voice. Notably, multiple candidate SVIs captured from different angles and timestamps might be retrieved for clips from the same recording. A sound source localization model \cite{park2024can} was further applied to match each clip with the most relevant landscape image. 
To ensure temporal consistency and account for environmental changes, we refined the dataset by discarding soundscape-SVI pairs with significant time gaps.
Furthermore, to obtain annotated place semantics, a Vision-Language Model (VLM), Qwen2.5-VL-7b, \cite{qwenvltechnicalreport2025} was used to annotate each pair with a scene context (e.g., street and residential neighborhood), which serve as the scene prompts for our model. In total, SoundingSVI dataset comprises 169,221 soundscape-landscape pairs in 90 countries.

\noindent\textbf{SonicUrban.}
The SonicUrban dataset leverages unedited videos as a source of temporally and spatially consistent soundscape-landscape pairs. 
Following a prior practice \cite{zhuang2024hearing}, we searched and manually checked YouTube videos that contain real-world soundscapes and landscapes, using a combination of city names and keywords, like ``City walk; New York''.
Each video was processed into 10-second audio clips, and 10 evenly distributed frames were extracted for each clip. The same human voice detection model \cite{silero2024vad} and sound source localization model \cite{park2024can} used in SoundingSVI were applied to filter human-voice dominated clips and identify the most representative frame for each audio clip, respectively. In total, the constructed SonicUrban consists of 236,674 soundscape-landscape pairs across 131 cities and 97 countries. The same VLM \cite{qwenvltechnicalreport2025} was further employed to generate a scene prompt for each pair as a geographic scene condition.
\section{Geo-Contextual Evaluation Suite}
Prior A2I studies commonly rely on evaluation metrics such as Fréchet Inception Distance (FID) \cite{heusel2017gans}, Audio-Image Similarity (AIS) \cite{heusel2017gans}, and Image-Image Similarity (IIS) \cite{biner2024sonicdiffusion}, to assess visual or audio fidelity. However, such metrics do not assess whether the generated image is coherent with the underlying geographic setting or the intended target scene. 
Since soundscapes and landscapes co-occur in the same space, they are expected to share similar environmental characteristics and place settings.
Following this hypothesis, we introduce \textbf{Place Similarity Score (PSS)}, a practically informed geo-contextual evaluation that quantifies the underlying place setting reflected in images across the element-, scene-, and human-perception levels.
These three levels of measurements of environmental characteristics have been widely employed in geographic and urban planning practices \cite{lynch1960image, kaplan1989experience}.

\noindent\textbf{Element-level PSS.}
This metric measures the agreement in geographic elements (e.g., trees \cite{li2015assessing}, sky \cite{gong2018mapping}, water bodies \cite{steele2014morphological}, traffic signs \cite{balali2015detection}, and buildings \cite{chen2022examining}) between the ground-truth and generated landscapes \cite{kang2020review}.
We use DeepLabV3 \cite{chen2017rethinking} pre-trained on ADE20K \cite{zhou2019semantic} to segment \(K\!=\!150\) predefined elements.
For each image \(i\), let \(\mathbf{e}_i\in\mathbb{R}^K\) and \(\hat{\mathbf{e}}_i\in\mathbb{R}^K\) be the normalized element-ratio vectors derived from the ground-truth image \(l_i\) and the generated image \(\hat{l}_i\), respectively. Formally, The element-level score is calculated as:
\begin{equation}
\mathrm{PSS}_{\text{elem}}
= \frac{1}{n}\sum_{i=1}^n 
\frac{\mathbf{e}_i^{\top}\,\hat{\mathbf{e}}_i}
{\|\mathbf{e}_i\|_2\,\|\hat{\mathbf{e}}_i\|_2},
\label{eq:PSS-elem}
\end{equation}
where \(\mathbf{e}_i=(e_{i1},\dots,e_{iK})^\top\) and \(\hat{\mathbf{e}}_i=(\hat e_{i1},\dots,\hat e_{iK})^\top\); \(n\) is the number of evaluation images.

\noindent\textbf{Scene-level PSS.} This metric is designed to evaluate the consistency of the entire environmental scene (e.g., forest, beach, or residential area) between the generated landscape images and their corresponding ground truth images \cite{kang2020review}. 
To infer scene categories, we employ a ResNet50 model \cite{he2016deep}, pre-trained on the Places365 dataset \cite{zhou2017places}, to identify 365 predefined scene categories.
For each soundscape \(s\), we predict the top-\(k\) scene labels
and evaluate whether the generated image \(\hat{l_i}\) aligns with the same scene category as its ground truth image \(l_i\).
\begin{equation}
\mathrm{PPS}_{\text{Scene}}
= \frac{1}{n} \sum_{i=1}^{n} \mathbf{1}\left\{P_i^{\text{k}} \cap T_i^{\text{k}}\neq \varnothing\right\},
\label{eq:PSS-Scene}
\end{equation}
where $k=1$ or $5$. $n$ is the total number of images, $P_i^{\text{k} } $ denotes the top-$k$ predicted category set for the $i$-th generated image, $T_i^{\text{k} } $ denotes the top-$k$ ground truth category set for the $i$-th real image.\\
\noindent\textbf{Human Perception-level PSS.}This metric is designed to evaluate whether the generated and real landscape images evoke similar human subjective perceptions and feelings,
such as whether a place makes people feel ``safe” or ``lively” \cite{zhang2018measuring}.
We utilize a DenseNet121 model \cite{huang2017densely}, pre-trained on the MIT Place Pulse dataset \cite{naik2014streetscore}, to measure six dimensions of human perceptions of environment, 
including safe, beautiful, depressing, lively, wealthy, and boring. For each image $i$, we compute six-dimensional place-perception scores for the ground-truth image $l_i$ and the generated image $\hat{l}_i$.
The perception-level PSS is defined as:
\begin{equation}
\mathrm{PSS}_{\text{perc}}
= \frac{1}{n}\sum_{i=1}^{n}
\left\| \mathbf{R}(l_i) - \mathbf{R}(\hat{l}_i) \right\|_{1},
\label{eq:pps-human}
\end{equation}
where $\mathbf{R}(\cdot)\in\mathbb{R}^{6}$ denotes the six-dimensional perception
score vector, and $n$ is the number of test images.

By integrating metrics across element-, scene-, and human perception levels, we provide a holistic geo-contextual evaluation framework. 
Informed by geographic domain knowledge and urban planning practices,
we extend beyond measuring visual quality to assessing whether generated landscape images are contextually aligned with the environmental characteristics in input soundscapes.

\begin{table*}
\centering
\footnotesize
    \caption{\textbf{Quantitative evaluation results across SounDiT and baseline models on two proposed datasets and benchmark datasets.} \(\dag\) indicates pre-trained models. We compare our SounDiT model with baselines using general metrics (FID, AIS, IIS) and our PSS metrics. 
    }
\resizebox{\textwidth}{!}{%
\begin{tabular}{c|c|ccc|ccc}
\toprule
 &   & \multicolumn{3}{c|}{\textbf{General Metrics}} & \multicolumn{3}{c}{\textbf{Place Similarity Score (PSS)}} \\
\midrule 
 Dataset&  Method& \textbf{FID}\(\downarrow\) & \textbf{AIS}\(\uparrow\)  & \textbf{IIS}\(\uparrow\) & \textbf{Element}\(\uparrow\) & \textbf{Scene}\(\uparrow\) & \textbf{Perception}\(\downarrow\)  \\
\midrule
\multicolumn{1}{c|}{
    \multirow[c]{9}{*}{ 
      \shortstack[c]{SoundingSVI\\(169K)}
    }
  } 
    & CoDi\(\dag\) \cite{tang2024any} & 87.027 & \textbf{0.641} & 0.612 & 0.400 & 0.302 & 0.800\\
    & Sound2Scene \cite{sung2023sound} & 78.137 & 0.586 & 0.623 & 0.467 & 0.374 & 0.735\\
    & AudioToken\(\dag\) \cite{yariv2023audiotoken} & 142.441 & 0.600 & 0.519 & 0.248 & 0.184 & 0.884\\										
    & AudioToken (SD1) \cite{yariv2023audiotoken} & 105.472 & 0.509 & 0.572 & 0.527 & 0.553 & 0.797\\	
    & AudioToken (SD2) \cite{yariv2023audiotoken} & 149.473 & 0.504 & 0.559 & 0.249 & 0.311 & 0.845\\	
    & GlueGen \cite{qin2023gluegen} & 74.250&0.500&0.566&0.400&0.461&0.776\\
    & PixArt + MHCA \cite{chen2023pixart} & 34.108&0.518&0.578&0.474&0.390&0.743\\
    & \cellcolor{lightblue}SounDiT (Ours) & \cellcolor{lightblue}\textbf{16.839} & \cellcolor{lightblue}0.538 & \cellcolor{lightblue}\textbf{0.753} & \cellcolor{lightblue}\textbf{0.572} &\cellcolor{lightblue} \textbf{0.753} & \cellcolor{lightblue}\textbf{0.729}\\
\midrule
\multicolumn{1}{c|}{
    \multirow[c]{9}{*}{ 
      \shortstack[c]{\shortstack{SonicUrban\\(237K)}}
    }
  } 
    & CoDi\(\dag\)  \cite{tang2024any} & 83.904&0.503&0.497&0.273&0.310&0.856 \\
    & Sound2Scene \cite{sung2023sound} &49.681&0.503&0.498&0.332&0.324&0.829\\
    & AudioToken\(\dag\) \cite{yariv2023audiotoken} & 127.337 & 0.497 & 0.501 & 0.203 & 0.199 & 0.976\\
    & AudioToken (SD1) \cite{yariv2023audiotoken} & 95.210&0.505&0.503&0.419&0.484&0.979\\
    & AudioToken (SD2) \cite{yariv2023audiotoken} & 165.014&0.502&0.504&0.298&0.455&0.980\\
    & GlueGen \cite{qin2023gluegen} &64.434&0.498&0.499&0.405&0.481&0.805\\
    & PixArt + MHCA \cite{chen2023pixart} & 41.456&0.517&0.592&0.411&0.396&0.796\\
    & \cellcolor{lightblue}SounDiT (Ours) & \cellcolor{lightblue}\textbf{11.553} & \cellcolor{lightblue}\textbf{0.520} & \cellcolor{lightblue}\textbf{0.706} & \cellcolor{lightblue}\textbf{0.520} &\cellcolor{lightblue} \textbf{0.739} & \cellcolor{lightblue}\textbf{0.759}\\
\bottomrule

\end{tabular}
}
    \label{table:result}
\end{table*}
\section{Experiment}
\subsection{Implementation Details}  
\noindent\textbf{Model Architecture and Training.}
We adopt the pre-trained Variational Autoencoder (VAE) modules from Latent Stable Diffusion \cite{rombach2022high} as our landscape encoder–decoder pair, which were originally trained on the COCO dataset \cite{lin2014microsoft}. 
For soundscape and scene prompt conditioning, we employ the ImageBind-Huge \cite{girdhar2023imagebind}, pre-trained on two million AudioSet clips \cite{gemmeke2017audio} covering a wide range of environmental, chosen for its accuracy--efficiency balance among several encoders\cite{wang2024omnibind}.
Our SounDiT is trained using a learning rate of $1\times10^{-4}$. Experiments are conducted on NVIDIA H100, A100 and A6000 GPUs. During training, we set the soundscape guidance scalar to $\mu=1.0$. For the scene condition, we employ a learnable scaling parameter $\alpha$ initialized to $0$ to stabilize early training by starting from an identity mapping. At inference stage, we apply classifier-free guidance with a scale of $4.0$ for both the soundscape and the scene prompt. 

\noindent\textbf{Baseline Models.} 
We compare against state of the art audio to image models: CoDi \cite{tang2024any}, Sound2Scene \cite{sung2023sound}, AudioToken \cite{yariv2023audiotoken}, GlueGen \cite{qin2023gluegen} and a PixArt \cite{chen2023pixart} variant augmented with multihead cross attention for the soundscape condition (PixArt+MHCA). For CoDi, we utilized its official pre-trained model. For AudioToken, we employed the pre-trained model without additional training, and trained it on our datasets using Stable Diffusion 1 and 2 as its backbone, respectively. The other baselines are training from scratch based on their released code and training settings. 


\noindent\textbf{Benchmark Evaluation Metrics.}
To evaluate the model performance, we adopted FID \cite{heusel2017gans}, AIS \cite{heusel2017gans}, and IIS \cite{biner2024sonicdiffusion}, effective in previous audio-to-image generation studies \cite{biner2024sonicdiffusion,sung2024soundbrush}.
FID compares Gaussian distributions of features extracted from real and generated images, using a pre-trained Inception model. 
AIS measures audio-image semantic similarity using Wav2CLIP \cite{wu2022wav2clip}. 
IIS evaluates image-image semantic similarity using CLIP-based embedding \cite{hessel2021clipscore}. 
\subsection{Main Results} 
\begin{figure*}
    \centering
    \includegraphics[width=\linewidth]{Figure/main_paper/main_result.jpg}
    \caption{\textbf{Visual comparison of landscape images} generated by baseline models (CoDi, Sound2Scene, AudioToken(AT), GlueGen, PixArt) and our proposed SounDiT across SoundingSVI and SonicUrban. Ground truth landscape images are provided in the last column. Our proposed SounDiT significant improvements over baseline models in generating geographical context.}
    \label{fig: Main_result}
\end{figure*}
Figure \ref{fig: Main_result} presents landscape images generated by our proposed SounDiT and baseline models across SoundingSVI and SonicUrban. The results generated by SounDiT show greater geo-contextual similarity to ground truth images,
highlighting potential applications of SounDiT in geography and urban planning. To comprehensively evaluate model performance, we adopt general metrics (FID, AIS, IIS) in addition to our proposed geo-contextual PSS framework.
As reported in Table \ref{table:result}, SounDiT consistently outperforms baselines on both SoundingSVI and SonicUrban datasets, with notable FID improvements (34\(\rightarrow\)16 and 41\(\rightarrow\)11), and achieves superior or competitive results across all metrics.
SounDiT consistently ranks highest under PSS, indicating its ability to ensure geographic coherence in generated images.
These results validate the effectiveness of PSS in measuring geo-contextual coherence. 
\begin{table}[!b]
\centering
\caption{\textbf{Component ablations on SoundingSVI (169K)} with two experts in the MoE Soundscape Conditioning module. We assess the individual contributions of \emph{Scene Low-Rank Content Mixer (SLRCM)} and the \emph{S-AdaLN.}}
\resizebox{\columnwidth}{!}{%
\begin{tabular}{lcccc}
\toprule
Variant & FID\(\downarrow\) &AIS\(\uparrow\) &IIS\(\uparrow\) & $\mathrm{PSS}_{\mathrm{Scene}}$\(\uparrow\)\\
\midrule
Full Model           & \textbf{19.195} &\textbf{0.538}& \textbf{0.750} & \textbf{0.734}\\
w/o SLRCM + S-AdaLN   & 25.375 &0.511& 0.539 & 0.428\\
w/o SLRCM            & 20.335 &0.534& 0.728 & 0.704\\
w/o S-AdaLN           & 23.435 &0.529& 0.629 & 0.572\\
\bottomrule
\end{tabular}
}
\label{table:ablation_scene}
\end{table}

\begin{figure}
    \centering
    \includegraphics[width=1\linewidth]{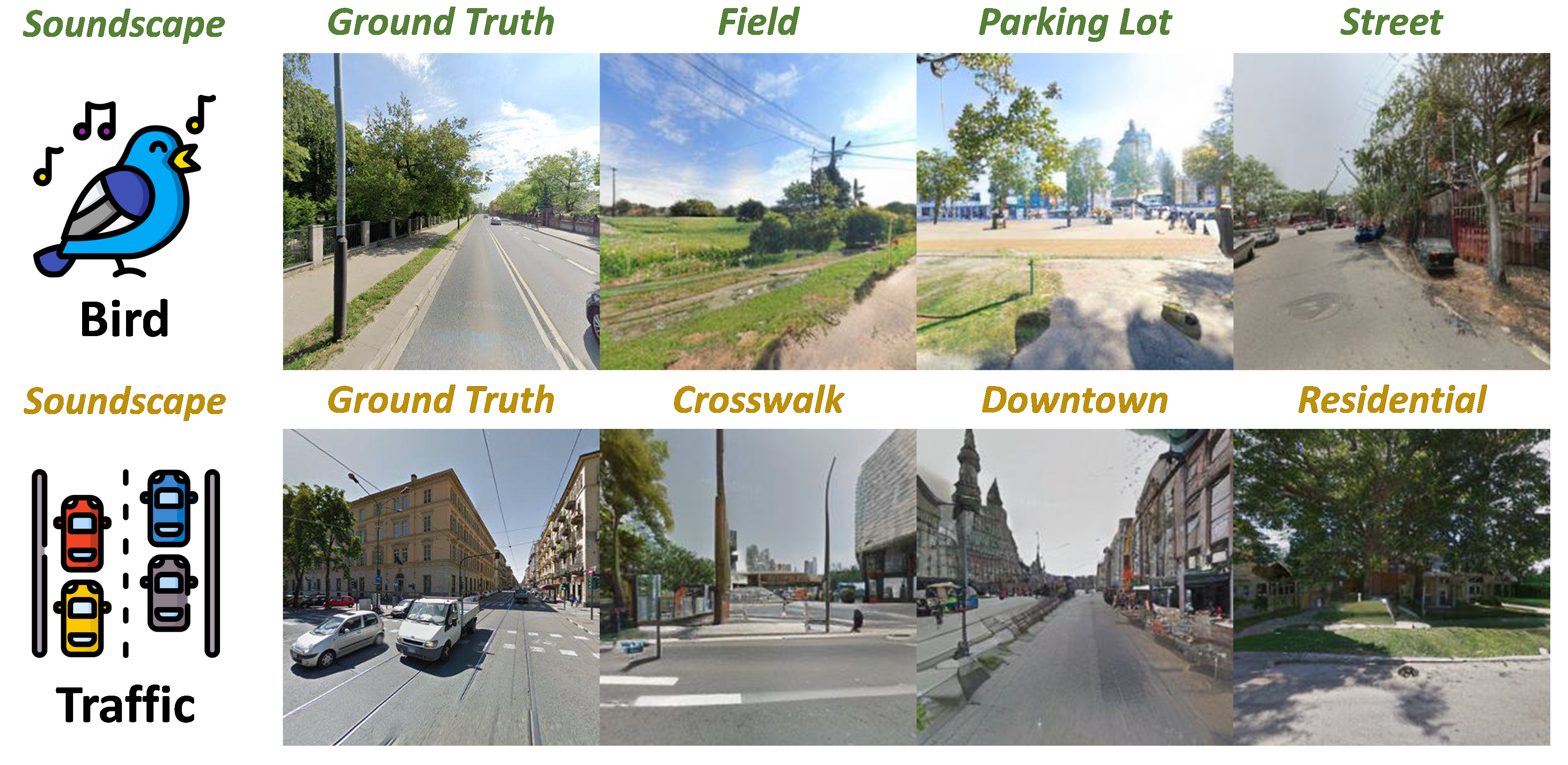}
    \caption{\textbf{Scene-conditioned generation.} With the same soundscape input, SounDiT generates visually distinct yet acoustically consistent images.}
    \vspace{-5mm}
    \label{fig:change_scene_results}
\end{figure}

\noindent\textbf{Practical Applications.}
To enhance real-world utility and user control, SounDiT supports scene-conditioned generation. Given a fixed soundscape, modifying the scene prompt generates geo-contextual, semantically coherent images that preserve both auditory information and scene context (Fig.~\ref{fig:change_scene_results}).
This supports practical applications such as soundscape-guided urban design, informing evidence-based design strategies to promote public health, safety, and environmental comfort. First, SounDiT can support soundscape-guided urban design by generating plausible visual place contexts from a target soundscape (quiet/restorative or a recorded clip), helping designers plan actionable elements such as greenery and water. Second, SounDiT facilities noise mitigation strategies by visualizing the built environment, enabling diagnosis of noise sources beyond decibel (dB), informing interventions (green buffers, traffic calming) for planning decisions. Moreover, SounDiT enables inclusive, accessible design by translating soundscape into visual landscapes, 
broadens participation in urban design\//planning practices, especially for hearing-impaired participants.

\noindent\textbf{User Study.} In addition to these quantitative metrics, we further
conducted a user study to assess whether the generated landscape images align with human perceptions of geographic contexts.
Specifically, 17 participants were asked to complete two matching tasks: (1) selecting the generated image that best aligned with the given soundscape; and (2) identifying which generated landscape image most closely matched a ground-truth landscape. The average matching accuracy is 86.13\%, indicating
a strong perceptual alignment between soundscapes and their corresponding generated images.
This demonstrates the effectiveness of SounDiT for GeoS2L.
\begin{table}[t]
\centering
\caption{\textbf{Expert scalability of MoE cross-attention on SoundingSVI (169K).} We vary the number of audio experts $M\!\in\!\{2,4,6,8\}$ while keeping other settings fixed. }

\begin{tabular}{lcccc}
\toprule
Experts & 2 & 4 & 6 & 8 \\
\midrule
FID\(\downarrow\)& 19.195& 18.304& 17.278& \textbf{16.839} \\
$\mathrm{PSS}_{\mathrm{Scene}}$\(\uparrow\) & 0.734&0.741& 0.742& \textbf{0.753} \\
\bottomrule
\end{tabular}

\label{table:ablation_expert}
\end{table}
\subsection{Ablations}

To validate our architectural design and quantify the impact of our proposed modules of SounDiT, we conduct two types of ablation experiments: component ablations and component scalability ablations. First, to assess the contribution of scene prompt and two scene conditioning modules, we perform experiments by first removing both SLRCM and S-AdaLN, and then removing each module individually. As shown in Table~\ref{table:ablation_scene}, removing either module results in a noticeable drop in performance. These results indicate that injecting scene condition both before and after the MoE Soundscape Conditioning module improves geo-contextual alignment and overall quality. Second, we test the scalability of MoE Soundscape Conditioning by varying the number of experts within this module (Table~\ref{table:ablation_expert}). Results show that the model with more experts achieves better performance in FID and scene consistency.
\section{Conclusion}
In this work, we present Geo-Contextual Soundscape-to-Landscape (GeoS2L), a practically-informed novel generation problem that synthesizes realistic landscape images from environmental soundscapes.
To support GeoS2L, we construct SoundingSVI and SonicUrban, two large-scale, multi-modal, geo-contextual datasets, capturing diverse sounding and visual environments.
We propose SounDiT, a scalable diffusion-based framework that captures both sounding environment and scene context by MoE Soundscape Conditioning and Scene Low-Rank Content Mixer.
We further design a novel geo-contextual evaluation framework, the Place Similarity Score (PSS), which assesses geographic and environmental consistency.
Extensive experiments demonstrate that SounDiT consistently outperforms existing baselines and can generate realistic landscape images.
The proposed geo-contextual datasets, model, and evaluation framework effectively integrate geographic knowledge into generative modeling, establishing a solid benchmark for the GeoS2L task. 



\section*{Acknowledgements}
We acknowledge ACCESS, supported by the U.S. National Science Foundation, for allocation SOC250002. We also acknowledge resources from the NAIRR Pilot (NAIRR250075). We acknowledge the Texas Advanced Computing Center (TACC) at The University of Texas at Austin for providing HPC resources.
{
    \small
    \bibliographystyle{ieeenat_fullname}
    \bibliography{main}

@String(ICASSP=	{ICASSP})

@inproceedings{peebles2023scalable,
  title={Scalable diffusion models with transformers},
  author={Peebles, William and Xie, Saining},
  booktitle={Proceedings of the IEEE/CVF International Conference on Computer Vision},
  pages={4195--4205},
  year={2023}
}

@inproceedings{lee2022sound,
  title={Sound-guided semantic video generation},
  author={Lee, Seung Hyun and Oh, Gyeongrok and Byeon, Wonmin and Kim, Chanyoung and Ryoo, Won Jeong and Yoon, Sang Ho and Cho, Hyunjun and Bae, Jihyun and Kim, Jinkyu and Kim, Sangpil},
  booktitle={European Conference on Computer Vision},
  pages={34--50},
  year={2022},
  organization={Springer}
}

@inproceedings{li2022learning,
  title={Learning visual styles from audio-visual associations},
  author={Li, Tingle and Liu, Yichen and Owens, Andrew and Zhao, Hang},
  booktitle={European Conference on Computer Vision},
  pages={235--252},
  year={2022},
  organization={Springer}
}

@inproceedings{sung2023sound,
  title={Sound to visual scene generation by audio-to-visual latent alignment},
  author={Sung-Bin, Kim and Senocak, Arda and Ha, Hyunwoo and Owens, Andrew and Oh, Tae-Hyun},
  booktitle={Proceedings of the IEEE/CVF Conference on Computer Vision and Pattern Recognition},
  pages={6430--6440},
  year={2023}
}

@article{goodfellow2014generative,
  title={Generative adversarial nets},
  author={Goodfellow, Ian and Pouget-Abadie, Jean and Mirza, Mehdi and Xu, Bing and Warde-Farley, David and Ozair, Sherjil and Courville, Aaron and Bengio, Yoshua},
  journal={Advances in neural information processing systems},
  volume={27},
  year={2014}
}

@inproceedings{rombach2022high,
  title={High-resolution image synthesis with latent diffusion models},
  author={Rombach, Robin and Blattmann, Andreas and Lorenz, Dominik and Esser, Patrick and Ommer, Bj{\"o}rn},
  booktitle={Proceedings of the IEEE/CVF conference on computer vision and pattern recognition},
  pages={10684--10695},
  year={2022}
}

@article{yang2024visual,
  title={Visual Echoes: A Simple Unified Transformer for Audio-Visual Generation},
  author={Yang, Shiqi and Zhong, Zhi and Zhao, Mengjie and Takahashi, Shusuke and Ishii, Masato and Shibuya, Takashi and Mitsufuji, Yuki},
  journal={arXiv preprint arXiv:2405.14598},
  year={2024}
}

@inproceedings{qin2023gluegen,
  title={Gluegen: Plug and play multi-modal encoders for x-to-image generation},
  author={Qin, Can and Yu, Ning and Xing, Chen and Zhang, Shu and Chen, Zeyuan and Ermon, Stefano and Fu, Yun and Xiong, Caiming and Xu, Ran},
  booktitle={Proceedings of the IEEE/CVF International Conference on Computer Vision},
  pages={23085--23096},
  year={2023}
}

@article{yariv2023audiotoken,
  title={Audiotoken: Adaptation of text-conditioned diffusion models for audio-to-image generation},
  author={Yariv, Guy and Gat, Itai and Wolf, Lior and Adi, Yossi and Schwartz, Idan},
  journal={arXiv preprint arXiv:2305.13050},
  year={2023}
}

@article{biner2024sonicdiffusion,
  title={SonicDiffusion: Audio-Driven Image Generation and Editing with Pretrained Diffusion Models},
  author={Biner, Burak Can and Sofian, Farrin Marouf and Karaka{\c{s}}, Umur Berkay and Ceylan, Duygu and Erdem, Erkut and Erdem, Aykut},
  journal={arXiv preprint arXiv:2405.00878},
  year={2024}
}

@article{tang2024any,
  title={Any-to-any generation via composable diffusion},
  author={Tang, Zineng and Yang, Ziyi and Zhu, Chenguang and Zeng, Michael and Bansal, Mohit},
  journal={Advances in Neural Information Processing Systems},
  volume={36},
  year={2024}
}

@inproceedings{girdhar2023imagebind,
  title={Imagebind: One embedding space to bind them all},
  author={Girdhar, Rohit and El-Nouby, Alaaeldin and Liu, Zhuang and Singh, Mannat and Alwala, Kalyan Vasudev and Joulin, Armand and Misra, Ishan},
  booktitle={Proceedings of the IEEE/CVF Conference on Computer Vision and Pattern Recognition},
  pages={15180--15190},
  year={2023}
}

@article{heidler2023self,
  title={Self-supervised audiovisual representation learning for remote sensing data},
  author={Heidler, Konrad and Mou, Lichao and Hu, Di and Jin, Pu and Li, Guangyao and Gan, Chuang and Wen, Ji-Rong and Zhu, Xiao Xiang},
  journal={International Journal of Applied Earth Observation and Geoinformation},
  volume={116},
  pages={103130},
  year={2023},
  publisher={Elsevier}
}

@article{li2024hunyuan,
  title={Hunyuan-DiT: A Powerful Multi-Resolution Diffusion Transformer with Fine-Grained Chinese Understanding},
  author={Li, Zhimin and Zhang, Jianwei and Lin, Qin and Xiong, Jiangfeng and Long, Yanxin and Deng, Xinchi and Zhang, Yingfang and Liu, Xingchao and Huang, Minbin and Xiao, Zedong and others},
  journal={arXiv preprint arXiv:2405.08748},
  year={2024}
}

@article{chen2023pixart,
  title={Pixart-$alpha$: Fast training of diffusion transformer for photorealistic text-to-image synthesis},
  author={Chen, Junsong and Yu, Jincheng and Ge, Chongjian and Yao, Lewei and Xie, Enze and Wu, Yue and Wang, Zhongdao and Kwok, James and Luo, Ping and Lu, Huchuan and others},
  journal={arXiv preprint arXiv:2310.00426},
  year={2023}
}

@article{zhang2018measuring,
  title={Measuring human perceptions of a large-scale urban region using machine learning},
  author={Zhang, Fan and Zhou, Bolei and Liu, Liu and Liu, Yu and Fung, Helene H and Lin, Hui and Ratti, Carlo},
  journal={Landscape and Urban Planning},
  volume={180},
  pages={148--160},
  year={2018},
  publisher={Elsevier}
}

@article{heusel2017gans,
  title={Gans trained by a two time-scale update rule converge to a local nash equilibrium},
  author={Heusel, Martin and Ramsauer, Hubert and Unterthiner, Thomas and Nessler, Bernhard and Hochreiter, Sepp},
  journal={Advances in neural information processing systems},
  volume={30},
  year={2017}
}

@article{zhou2019semantic,
  title={Semantic understanding of scenes through the ade20k dataset},
  author={Zhou, Bolei and Zhao, Hang and Puig, Xavier and Xiao, Tete and Fidler, Sanja and Barriuso, Adela and Torralba, Antonio},
  journal={International Journal of Computer Vision},
  volume={127},
  number={3},
  pages={302--321},
  year={2019},
  publisher={Springer}
}

@article{zhou2017places,
   title={Places: A 10 million Image Database for Scene Recognition},
   author={Zhou, Bolei and Lapedriza, Agata and Khosla, Aditya and Oliva, Aude and Torralba, Antonio},
   journal={IEEE Transactions on Pattern Analysis and Machine Intelligence},
   year={2017},
   publisher={IEEE}
 }

@inproceedings{chen2020vggsound,
  title={Vggsound: A large-scale audio-visual dataset},
  author={Chen, Honglie and Xie, Weidi and Vedaldi, Andrea and Zisserman, Andrew},
  booktitle={ICASSP 2020-2020 IEEE International Conference on Acoustics, Speech and Signal Processing (ICASSP)},
  pages={721--725},
  year={2020},
  organization={IEEE}
}

@inproceedings{zhou2018visual,
  title={Visual to sound: Generating natural sound for videos in the wild},
  author={Zhou, Yipin and Wang, Zhaowen and Fang, Chen and Bui, Trung and Berg, Tamara L},
  booktitle={Proceedings of the IEEE conference on computer vision and pattern recognition},
  pages={3550--3558},
  year={2018}
}

@inproceedings{owens2016visually,
  title={Visually indicated sounds},
  author={Owens, Andrew and Isola, Phillip and McDermott, Josh and Torralba, Antonio and Adelson, Edward H and Freeman, William T},
  booktitle={Proceedings of the IEEE conference on computer vision and pattern recognition},
  pages={2405--2413},
  year={2016}
}

@article{livingstone2018ryerson,
  title={The Ryerson Audio-Visual Database of Emotional Speech and Song (RAVDESS): A dynamic, multimodal set of facial and vocal expressions in North American English},
  author={Livingstone, Steven R and Russo, Frank A},
  journal={PloS one},
  volume={13},
  number={5},
  pages={e0196391},
  year={2018},
  publisher={Public Library of Science San Francisco, CA USA}
}

@article{nagrani2017voxceleb,
  title={Voxceleb: a large-scale speaker identification dataset},
  author={Nagrani, Arsha and Chung, Joon Son and Zisserman, Andrew},
  journal={arXiv preprint arXiv:1706.08612},
  year={2017}
}

@inproceedings{yuan2012discovering,
  title={Discovering regions of different functions in a city using human mobility and POIs},
  author={Yuan, Jing and Zheng, Yu and Xie, Xing},
  booktitle={Proceedings of the 18th ACM SIGKDD international conference on Knowledge discovery and data mining},
  pages={186--194},
  year={2012}
}

@inproceedings{chen2024pixart,
  title={Pixart-$\sigma$: Weak-to-strong training of diffusion transformer for 4k text-to-image generation},
  author={Chen, Junsong and Ge, Chongjian and Xie, Enze and Wu, Yue and Yao, Lewei and Ren, Xiaozhe and Wang, Zhongdao and Luo, Ping and Lu, Huchuan and Li, Zhenguo},
  booktitle={European Conference on Computer Vision},
  pages={74--91},
  year={2024},
  organization={Springer}
}

@article{ma2024latte,
  title={Latte: Latent diffusion transformer for video generation},
  author={Ma, Xin and Wang, Yaohui and Jia, Gengyun and Chen, Xinyuan and Liu, Ziwei and Li, Yuan-Fang and Chen, Cunjian and Qiao, Yu},
  journal={arXiv preprint arXiv:2401.03048},
  year={2024}
}

@article{wang2024av,
  title={AV-DiT: Efficient Audio-Visual Diffusion Transformer for Joint Audio and Video Generation},
  author={Wang, Kai and Deng, Shijian and Shi, Jing and Hatzinakos, Dimitrios and Tian, Yapeng},
  journal={arXiv preprint arXiv:2406.07686},
  year={2024}
}

@article{liu2024sora,
  title={Sora: A review on background, technology, limitations, and opportunities of large vision models},
  author={Liu, Yixin and Zhang, Kai and Li, Yuan and Yan, Zhiling and Gao, Chujie and Chen, Ruoxi and Yuan, Zhengqing and Huang, Yue and Sun, Hanchi and Gao, Jianfeng and others},
  journal={arXiv preprint arXiv:2402.17177},
  year={2024}
}

@inproceedings{lee2021acav100m,
  title={Acav100m: Automatic curation of large-scale datasets for audio-visual video representation learning},
  author={Lee, Sangho and Chung, Jiwan and Yu, Youngjae and Kim, Gunhee and Breuel, Thomas and Chechik, Gal and Song, Yale},
  booktitle={Proceedings of the IEEE/CVF International Conference on Computer Vision},
  pages={10274--10284},
  year={2021}
}

@inproceedings{park2024can,
  title={Can clip help sound source localization?},
  author={Park, Sooyoung and Senocak, Arda and Chung, Joon Son},
  booktitle={Proceedings of the IEEE/CVF Winter Conference on Applications of Computer Vision},
  pages={5711--5720},
  year={2024}
}

@article{noll2013radio,
  title={radio aporee},
  author={Noll, Udo},
  journal={Berlin: Udo Noll},
  year={2013}
}

@article{sung2024soundbrush,
  title={SoundBrush: Sound as a Brush for Visual Scene Editing},
  author={Sung-Bin, Kim and Jun-Seong, Kim and Ko, Junseok and Kim, Yewon and Oh, Tae-Hyun},
  journal={arXiv preprint arXiv:2501.00645},
  year={2024}
}

@article{chen2017rethinking,
  title={Rethinking atrous convolution for semantic image segmentation},
  author={Chen, Liang-Chieh},
  journal={arXiv preprint arXiv:1706.05587},
  year={2017}
}

@inproceedings{naik2014streetscore,
  title={Streetscore-predicting the perceived safety of one million streetscapes},
  author={Naik, Nikhil and Philipoom, Jade and Raskar, Ramesh and Hidalgo, C{\'e}sar},
  booktitle={Proceedings of the IEEE conference on computer vision and pattern recognition workshops},
  pages={779--785},
  year={2014}
}

@book{ruiz2024urban,
  title={Urban Soundscapes: A Guide to Listening for Landscape Architecture and Urban Design},
  author={Ruiz, Arana and others},
  year={2024},
  publisher={Newcastle University}
}

@book{berleant1997living,
  title={Living in the landscape: Toward an aesthetics of environment},
  author={Berleant, Arnold},
  year={1997},
  publisher={University press of Kansas}
}

@article{blesser2007spaces,
  title={Spaces speak, are you listening},
  author={Blesser, Barry and Salter, Linda-Ruth},
  journal={Experiencing aural architecture},
  volume={232},
  year={2007},
  publisher={MIT press Cambridge, MA, USA}
}

@article{algargoosh2022impact,
  title={The impact of the acoustic environment on human emotion and experience: A case study of worship spaces},
  author={Algargoosh, Alaa and Soleimani, Babak and O’Modhrain, Sile and Navvab, Mojtaba},
  journal={Building Acoustics},
  volume={29},
  number={1},
  pages={85--106},
  year={2022},
  publisher={SAGE Publications Sage UK: London, England}
}

@article{tuan1975images,
  title={Images and mental maps},
  author={Tuan, Yi-fu},
  journal={Annals of the Association of American geographers},
  volume={65},
  number={2},
  pages={205--212},
  year={1975},
  publisher={Wiley Online Library}
}

@book{cosgrove2012geography,
  title={Geography and vision},
  author={Cosgrove, Denis},
  year={2012},
  publisher={IB Tauris}
}

@incollection{krygier1994sound,
  title={Sound and geographic visualization},
  author={Krygier, John B},
  booktitle={Modern cartography series},
  volume={2},
  pages={149--166},
  year={1994},
  publisher={Elsevier}
}

@inproceedings{de2010soundscape,
  title={The soundscape approach for early stage urban planning: a case study},
  author={De Coensel, Bert and Bockstael, Annelies and Dekoninck, Luc and Botteldooren, Dick and Schulte-Fortkamp, Brigitte and Kang, Jian and Nilsson, Mats E},
  booktitle={INTER-NOISE and NOISE-CON Congress and Conference Proceedings},
  volume={2010},
  number={8},
  pages={3294--3303},
  year={2010},
  organization={Institute of Noise Control Engineering}
}

@article{huang2024estimating,
  title={Estimating urban noise along road network from street view imagery},
  author={Huang, Jing and Fei, Teng and Kang, Yuhao and Li, Jun and Liu, Ziyu and Wu, Guofeng},
  journal={International Journal of Geographical Information Science},
  volume={38},
  number={1},
  pages={128--155},
  year={2024},
  publisher={Taylor \& Francis}
}

@book{mehrabian1974approach,
  title={An approach to environmental psychology.},
  author={Mehrabian, Albert and Russell, James A},
  year={1974},
  publisher={the MIT Press}
}

@article{guastavino2007categorization,
  title={Categorization of environmental sounds.},
  author={Guastavino, Catherine},
  journal={Canadian Journal of Experimental Psychology/Revue canadienne de psychologie exp{\'e}rimentale},
  volume={61},
  number={1},
  pages={54},
  year={2007},
  publisher={Canadian Psychological Association}
}

@article{wang2024omnibind,
  title={Omnibind: Large-scale omni multimodal representation via binding spaces},
  author={Wang, Zehan and Zhang, Ziang and Zhang, Hang and Liu, Luping and Huang, Rongjie and Cheng, Xize and Zhao, Hengshuang and Zhao, Zhou},
  journal={arXiv preprint arXiv:2407.11895},
  year={2024}
}

@inproceedings{gemmeke2017audio,
  title={Audio set: An ontology and human-labeled dataset for audio events},
  author={Gemmeke, Jort F and Ellis, Daniel PW and Freedman, Dylan and Jansen, Aren and Lawrence, Wade and Moore, R Channing and Plakal, Manoj and Ritter, Marvin},
  booktitle={IEEE international conference on acoustics, speech and signal processing (ICASSP)},
  pages={776--780},
  year={2017},
  organization={IEEE}
}

@inproceedings{wu2022wav2clip,
    title={Wav2CLIP: Learning Robust Audio Representations From CLIP},
    author={Wu, Ho-Hsiang and Seetharaman, Prem and Kumar, Kundan and Bello, Juan Pablo},
    booktitle={ICASSP 2022 - 2022 IEEE International Conference on Acoustics, Speech and Signal Processing (ICASSP)},
    year={2022}
}

@article{kang2020review,
  title={A review of urban physical environment sensing using street view imagery in public health studies},
  author={Kang, Yuhao and Zhang, Fan and Gao, Song and Lin, Hui and Liu, Yu},
  journal={Annals of GIS},
  volume={26},
  number={3},
  pages={261--275},
  year={2020},
  publisher={Taylor \& Francis}
}

@article{li2015assessing,
  title={Assessing street-level urban greenery using Google Street View and a modified green view index},
  author={Li, Xiaojiang and Zhang, Chuanrong and Li, Weidong and Ricard, Robert and Meng, Qingyan and Zhang, Weixing},
  journal={Urban Forestry \& Urban Greening},
  volume={14},
  number={3},
  pages={675--685},
  year={2015},
  publisher={Elsevier}
}

@article{gong2018mapping,
  title={Mapping sky, tree, and building view factors of street canyons in a high-density urban environment},
  author={Gong, Fang-Ying and Zeng, Zhao-Cheng and Zhang, Fan and Li, Xiaojiang and Ng, Edward and Norford, Leslie K},
  journal={Building and Environment},
  volume={134},
  pages={155--167},
  year={2018},
  publisher={Elsevier}
}

@inproceedings{he2016deep,
  title={Deep residual learning for image recognition},
  author={He, Kaiming and Zhang, Xiangyu and Ren, Shaoqing and Sun, Jian},
  booktitle={Proceedings of the IEEE conference on computer vision and pattern recognition},
  pages={770--778},
  year={2016}
}

@article{kong2024hunyuanvideo,
  title={Hunyuanvideo: A systematic framework for large video generative models},
  author={Kong, Weijie and Tian, Qi and Zhang, Zijian and Min, Rox and Dai, Zuozhuo and Zhou, Jin and Xiong, Jiangfeng and Li, Xin and Wu, Bo and Zhang, Jianwei and others},
  journal={arXiv preprint arXiv:2412.03603},
  year={2024}
}

@inproceedings{lin2014microsoft,
  title={Microsoft coco: Common objects in context},
  author={Lin, Tsung-Yi and Maire, Michael and Belongie, Serge and Hays, James and Perona, Pietro and Ramanan, Deva and Doll{\'a}r, Piotr and Zitnick, C Lawrence},
  booktitle={Computer vision--ECCV 2014: 13th European conference, zurich, Switzerland, September 6-12, 2014, proceedings, part v 13},
  pages={740--755},
  year={2014},
  organization={Springer}
}

@article{zhuang2024hearing,
  title={From hearing to seeing: Linking auditory and visual place perceptions with soundscape-to-image generative artificial intelligence},
  author={Zhuang, Yonggai and Kang, Yuhao and Fei, Teng and Bian, Meng and Du, Yunyan},
  journal={Computers, Environment and Urban Systems},
  volume={110},
  pages={102122},
  year={2024},
  publisher={Elsevier}
}

@article{fang2021soundscape,
  title={Soundscape perceptions and preferences for different groups of users in urban recreational forest parks},
  author={Fang, Xingyue and Gao, Tian and Hedblom, Marcus and Xu, Naisheng and Xiang, Yi and Hu, Mengyao and Chen, Yuxuan and Qiu, Ling},
  journal={Forests},
  volume={12},
  number={4},
  pages={468},
  year={2021},
  publisher={MDPI}
}

@inproceedings{kahl-2019-overview,
  author    = {Stefan Kahl and
               Fabian{-}Robert St{\"{o}}ter and
               Herv{\'{e}} Go{\"{e}}au and
               Herv{\'{e}} Glotin and
               Bob Planqu{\'{e}} and
               Willem{-}Pier Vellinga and
               Alexis Joly},
  editor    = {Linda Cappellato and
               Nicola Ferro and
               David E. Losada and
               Henning M{\"{u}}ller},
  title     = {Overview of BirdCLEF 2019: Large-Scale Bird Recognition in Soundscapes},
  booktitle = {Working Notes of {CLEF} 2019 - Conference and Labs of the Evaluation
               Forum, Lugano, Switzerland, September 9-12, 2019},
  series    = {{CEUR} Workshop Proceedings},
  volume    = {2380},
  publisher = {CEUR-WS.org},
  year      = {2019},
  bibsource = {dblp computer science bibliography, https://dblp.org}
}

@article{rauch_birdset_2024,
	title = {{BirdSet}: A Large-Scale Dataset for Audio Classification in Avian Bioacoustics},
	doi = {10.48550/arXiv.2403.10380},
	shorttitle = {{BirdSet}},
	pages = {arXiv:2403.10380},
    journal = {{arXiv} e-prints},
	journaltitle = {{arXiv} e-prints},
	author = {Rauch, Lukas and Schwinger, Raphael and Wirth, Moritz and Heinrich, René and Huseljic, Denis and Herde, Marek and Lange, Jonas and Kahl, Stefan and Sick, Bernhard and Tomforde, Sven and Scholz, Christoph},
	urldate = {2025-04-15},
	date = {2024-03},
	langid = {english},
    year = {2024},
}

@inproceedings{gemmeke_audio_2017,
	title = {Audio Set: An ontology and human-labeled dataset for audio events},
	doi = {10.1109/ICASSP.2017.7952261},
	shorttitle = {Audio Set},
	eventtitle = {2017 {IEEE} International Conference on Acoustics, Speech and Signal Processing ({ICASSP})},
	pages = {776--780},
	booktitle = {2017 {IEEE} International Conference on Acoustics, Speech and Signal Processing ({ICASSP})},
	author = {Gemmeke, Jort F. and Ellis, Daniel P. W. and Freedman, Dylan and Jansen, Aren and Lawrence, Wade and Moore, R. Channing and Plakal, Manoj and Ritter, Marvin},
	urldate = {2025-04-15},
	date = {2017-03},
	note = {{ISSN}: 2379-190X},
    year = {2017},
}

@inproceedings{salamon_dataset_2014,
	location = {New York, {NY}, {USA}},
	title = {A Dataset and Taxonomy for Urban Sound Research},
	isbn = {978-1-4503-3063-3},
	doi = {10.1145/2647868.2655045},
     year = {2014},
	series = {{MM} '14},
	pages = {1041--1044},
	booktitle = {Proceedings of the 22nd {ACM} international conference on Multimedia},
	publisher = {Association for Computing Machinery},
	author = {Salamon, Justin and Jacoby, Christopher and Bello, Juan Pablo},
	urldate = {2025-04-15},
	date = {2014-11-03},
}

@inproceedings{cartwright_sonyc-ust-v2_2020,
    Author = "Cartwright, Mark and Mendez, Ana Elisa Mendez and Cramer, Jason and Lostanlen, Vincent and Dove, Graham and Wu, Ho-Hsiang and Salamon, Justin and Nov, Oded and Bello, Juan",
    title = "{SONYC} Urban Sound Tagging ({SONYC-UST}): A Multilabel Dataset from an Urban Acoustic Sensor Network",
    year = "2019",
    booktitle = "Proceedings of the Workshop on Detection and Classification of Acoustic Scenes and Events (DCASE)",
    month = "October",
    pages = "35--39",
    url = "http://dcase.community/documents/workshop2019/proceedings/DCASE2019Workshop\_Cartwright\_4.pdf"
}

@article{green_eigenscape_2017,
	title = {{EigenScape}: A Database of Spatial Acoustic Scene Recordings},
	volume = {7},
	rights = {http://creativecommons.org/licenses/by/3.0/},
	issn = {2076-3417},
	doi = {10.3390/app7111204},
	shorttitle = {{EigenScape}},
	pages = {1204},
	number = {11},
    journal = {Applied Sciences},
    year = {2017},
	journaltitle = {Applied Sciences},
	author = {Green, Marc Ciufo and Murphy, Damian},
	urldate = {2025-04-15},
	date = {2017-11},
	langid = {english},
	note = {Number: 11
Publisher: Multidisciplinary Digital Publishing Institute},
}

@inproceedings{fan_emo-soundscapes_2017,
    year = {2017},
	title = {Emo-soundscapes: A dataset for soundscape emotion recognition},
	doi = {10.1109/ACII.2017.8273600},
	shorttitle = {Emo-soundscapes},
	eventtitle = {2017 Seventh International Conference on Affective Computing and Intelligent Interaction ({ACII})},
	pages = {196--201},
	booktitle = {2017 Seventh International Conference on Affective Computing and Intelligent Interaction ({ACII})},
	author = {Fan, Jianyu and Thorogood, Miles and Pasquier, Philippe},
	urldate = {2025-04-15},
	date = {2017-10},
	note = {{ISSN}: 2156-8111},
}

@article{liu_landscape_2013,
	title = {Landscape effects on soundscape experience in city parks},
	volume = {454-455},
	issn = {0048-9697},
    year = {2013},
	doi = {10.1016/j.scitotenv.2013.03.038},
	pages = {474--481},
    journal = {Science of The Total Environment},
	journaltitle = {Science of The Total Environment},
	shortjournal = {Science of The Total Environment},
	author = {Liu, Jiang and Kang, Jian and Luo, Tao and Behm, Holger},
	urldate = {2025-04-15},
	date = {2013-06-01},
}

@article{fuller_connecting_2015,
	title = {Connecting soundscape to landscape: Which acoustic index best describes landscape configuration?},
	volume = {58},
    year = {2015},
	issn = {1470-160X},
	doi = {10.1016/j.ecolind.2015.05.057},
	shorttitle = {Connecting soundscape to landscape},
	pages = {207--215},
    journal = {Ecological Indicators},
	journaltitle = {Ecological Indicators},
	shortjournal = {Ecological Indicators},
	author = {Fuller, Susan and Axel, Anne C. and Tucker, David and Gage, Stuart H.},
	urldate = {2025-04-15},
	date = {2015-11-01},
}

@book{Schafer1977-SCHTTO-12,
	address = {Philadelphia},
	author = {R. Murray Schafer},
	editor = {},
	publisher = {University of Pennsylvania Press},
	title = {The Tuning of the World: Toward a Theory of Soundscape Design},
	year = {1977}
}

@book{10.7551/mitpress/9780262015523.001.0001,
    author = {Malpas, Jeff},
    title = {The Place of Landscape: Concepts, Contexts, Studies},
    publisher = {The MIT Press},
    year = {2011},
    month = {05},
    isbn = {9780262295840},
    doi = {10.7551/mitpress/9780262015523.001.0001},
}

@article{chen2022examining,
  title={Examining the association between the built environment and pedestrian volume using street view images},
  author={Chen, Long and Lu, Yi and Ye, Yu and Xiao, Yang and Yang, Linchuan},
  journal={Cities},
  volume={127},
  pages={103734},
  year={2022},
  publisher={Elsevier}
}

@article{balali2015detection,
  title={Detection, classification, and mapping of US traffic signs using google street view images for roadway inventory management},
  author={Balali, Vahid and Ashouri Rad, Armin and Golparvar-Fard, Mani},
  journal={Visualization in Engineering},
  volume={3},
  pages={1--18},
  year={2015},
  publisher={Springer}
}

@article{steele2014morphological,
  title={Morphological characteristics of urban water bodies: mechanisms of change and implications for ecosystem function},
  author={Steele, MK and Heffernan, JB},
  journal={Ecological Applications},
  volume={24},
  number={5},
  pages={1070--1084},
  year={2014},
  publisher={Wiley Online Library}
}

@inproceedings{huang2017densely,
  title={Densely connected convolutional networks},
  author={Huang, Gao and Liu, Zhuang and Van Der Maaten, Laurens and Weinberger, Kilian Q},
  booktitle={Proceedings of the IEEE conference on computer vision and pattern recognition},
  pages={4700--4708},
  year={2017}
}

@misc{silero2024vad,
  author       = {Silero Team},
  title        = {Silero VAD: pre-trained enterprise-grade Voice Activity Detector (VAD), Number Detector and Language Classifier},
  year         = {2024},
}

@article{hessel2021clipscore,
  title={Clipscore: A reference-free evaluation metric for image captioning},
  author={Hessel, Jack and Holtzman, Ari and Forbes, Maxwell and Bras, Ronan Le and Choi, Yejin},
  journal={arXiv preprint arXiv:2104.08718},
  year={2021}
}

@article{lynch1960image,
  title={The image of the environment},
  author={Lynch, Kevin},
  journal={The image of the city},
  volume={11},
  pages={1--13},
  year={1960}
}

@book{kaplan1989experience,
  title={The experience of nature: A psychological perspective},
  author={Kaplan, Rachel and Kaplan, Stephen},
  year={1989},
  publisher={Cambridge university press}
}

@article{rasmussen2024sound,
  title={Sound evidence for biodiversity monitoring},
  author={Rasmussen, Jeppe H and Stowell, Dan and Briefer, Elodie F},
  journal={Science},
  volume={385},
  number={6705},
  pages={138--140},
  year={2024},
  publisher={American Association for the Advancement of Science}
}

@article{zhuang2025soundscape,
  title={Soundscape-to-panorama: spatialize auditory perception by linking acoustic environment to panorama},
  author={Zhuang, Yonggai and Gui, Yanni and Fei, Teng},
  journal={International Journal of Digital Earth},
  volume={18},
  number={1},
  pages={2545584},
  year={2025},
  publisher={Taylor \& Francis}
}

@article{kang2023soundscape,
  title={Soundscape in city and built environment: current developments and design potentials},
  author={Kang, Jian},
  journal={City and Built Environment},
  volume={1},
  number={1},
  pages={1},
  year={2023},
  publisher={Springer}
}

@article{hemmat2023exploring,
  title={Exploring noise pollution, causes, effects, and mitigation strategies: a review paper},
  author={Hemmat, Walihabib and Hesam, Atiq Mohammad and Atifnigar, Hamza},
  journal={European Journal of Theoretical and Applied Sciences},
  volume={1},
  number={5},
  pages={995--1005},
  year={2023}
}

@techreport{qwenvltechnicalreport2025,
    title = {Qwen2. 5-vl technical report},
    author = {Shuai Bai},
    institution = {Alibaba Damo Academy},
    year = {2025}
}

@article{goyal2017accurate,
  title={Accurate, large minibatch sgd: Training imagenet in 1 hour},
  author={Goyal, Priya and Doll{\'a}r, Piotr and Girshick, Ross and Noordhuis, Pieter and Wesolowski, Lukasz and Kyrola, Aapo and Tulloch, Andrew and Jia, Yangqing and He, Kaiming},
  journal={arXiv preprint arXiv:1706.02677},
  year={2017}
}

@article{uebel2021urban,
  title={Urban green space soundscapes and their perceived restorativeness},
  author={Uebel, Konrad and Marselle, Melissa and Dean, Angela J and Rhodes, Jonathan R and Bonn, Aletta},
  journal={People and Nature},
  volume={3},
  number={3},
  pages={756--769},
  year={2021},
  publisher={Wiley Online Library}
}

@article{chen2023natural,
  title={Natural sounds can encourage social interactions in urban parks},
  author={Chen, Xiaochao and Kang, Jian},
  journal={Landscape and Urban Planning},
  volume={239},
  pages={104870},
  year={2023},
  publisher={Elsevier}
}

@article{brown2004approach,
  title={An approach to the acoustic design of outdoor space},
  author={Brown, AL and Muhar, Andreas},
  journal={Journal of Environmental planning and Management},
  volume={47},
  number={6},
  pages={827--842},
  year={2004},
  publisher={Taylor \& Francis}
}

@inproceedings{brown2014soundscape,
  title={Soundscape planning as a complement to environmental noise management},
  author={Brown, Alan Lex},
  booktitle={InterNoise14: Improving the World through Noise Control},
  year={2014},
  organization={Australian Acoustical Society}
}

@article{jeon2012acoustical,
  title={Acoustical characteristics of water sounds for soundscape enhancement in urban open spaces},
  author={Jeon, Jin Yong and Lee, Pyoung Jik and You, Jin and Kang, Jian},
  journal={The Journal of the Acoustical Society of America},
  volume={131},
  number={3},
  pages={2101--2109},
  year={2012},
  publisher={AIP Publishing}
}

@article{zhang2017effects,
  title={Effects of soundscape on the environmental restoration in urban natural environments},
  author={Zhang, Yuan and Kang, Jian and Kang, Joe},
  journal={Noise and Health},
  volume={19},
  number={87},
  pages={65--72},
  year={2017},
  publisher={Medknow}
}

@article{jo2021urban,
  title={Urban soundscape categorization based on individual recognition, perception, and assessment of sound environments},
  author={Jo, Hyun In and Jeon, Jin Yong},
  journal={Landscape and urban planning},
  volume={216},
  pages={104241},
  year={2021},
  publisher={Elsevier}
}
}


\end{document}